\documentclass[sn-mathphys]{sn-jnl}
\begin{document}
	\title[Superposition in Measuring Apparatus]{Superposition in Measuring Apparatus: A Thought Experiment}
	\author*{\fnm{Vishwas} \sur{KS}}\email{vishwasks@iitb.ac.in}
	\abstract{The measurement problem in quantum mechanics arises from the apparent
		collapse of a superposition state to a definite outcome when a measurement
		is made. Although treating the measuring apparatus as a classical
		system has been a successful approach in explaining quantum phenomena,
		it raises fundamental questions about the nature of measurement and
		the validity of wave function collapse. In this paper, we discuss
		a thought experiment that explores superposition in the measuring
		apparatus when it is treated as a quantum system. The experiment uses
		the Hong-Ou-Mandel effect in a two-photon interference setup, and
		its outcome is indicated by the coincidence count. Specifically, a
		zero count implies the existence of superposition, while a non-zero
		count indicates a wave function collapse. The discussions provide
		insight into the measurement problem, particularly regarding wave
		function collapse and nested measurement, and highlight the importance
		of indistinguishability to it. It provides a framework that probes
		the exact conditions necessary for a wave function collapse to happen.}
	
	
	\maketitle

\section{Introduction}

The measurement problem has been a challenge in quantum mechanics
since its early days. According to quantum mechanics, an isolated
quantum system evolves over time according to Schr{\"o}dinger's equation.
However, when a measuring apparatus is used to observe the system,
the state of the system collapses probabilistically to one of the
eigenstates that form the superposition state. This process, known
as the collapse of the wave function, is a fundamental aspect of traditional
quantum mechanics \cite{von2018mathematical,omnes1999understanding,landsman2017foundations}.
In this context, the measuring apparatus is typically treated as a
classical system that cannot exist in any superposition state. This
approach has been highly successful in explaining experimental outcomes
of quantum phenomena in our classical macroworld.

If the measuring apparatus is treated as a quantum system, rather
than a classical one, then the act of measurement becomes an interaction
between two quantum systems. According to quantum mechanics, these
two systems would become entangled with each other, forming an entangled
composite system \cite{von2018mathematical,landsman2017foundations}.
This raises important questions about what exactly constitutes a measurement,
how it differs from other interactions between two quantum systems,
and whether wave function collapse is a real physical phenomenon.
These questions collectively form what is known as the measurement
problem \cite{zeilinger1996interpretation,landsman2017foundations,von2018mathematical}.
They have profound implications for our understanding of fundamental
reality and are still being debated among physicists today.

The aim of this paper is to present and discuss a thought experiment
in which a measuring apparatus could exhibit a superposition state.
In the thought experiment, this is accomplished by passing on the
superposition state from the measuring apparatus to a photon and performing
a two-photon interference. This thought experiment may help us understand
wave function collapse and nested measurement better (nested measurement
is when one performs measurements on a measuring apparatus).

The earliest formal discussions of measuring apparatus existing in
a superposition state were probably by von Neumann. He concluded that
treating all measuring apparatuses as quantum systems would lead to
an infinite regress, where superposition never collapses \cite{von2018mathematical}.
Schr{\"o}dinger discussed that the Geiger counter that measures radioactivity
in his Schr{\"o}dinger's cat thought experiment \cite{trimmer1980proceedings}
would be in a superposition state after measurement. Wigner similarly
discussed the possibility of the friend in his Wigner's friend thought
experiment \cite{wigner1961scientist} existing in a superposition
with respect to the outside observer. Experiments have now used the
measuring apparatus existing in a superposition state to understand
measurement and wave function collapse better \cite{carvalho2020decay,brune1996observing}.

Regarding wave function collapse, rather than treating the measuring
apparatus as a classical system, different interpretations of quantum
mechanics propose alternative ideas. Everett's relative state \cite{everett1957relative}
formulation eliminates the collapse postulate and derives experimental
outcomes identical to those obtained by assuming the collapse postulate,
thus also providing a justification for its use. Decoherence theory
\cite{zurek1991decoherence} explains why it is appropriate to treat
the measuring apparatus as classical, concluding that quantum effects
would be vanishingly small to observe in our everyday macroworld.
Objective collapse theories add new terms to Schr{\"o}dinger's equation
in order to account for the phenomenon of wave function collapse \cite{ghirardi1986unified,ghirardi1990continuous,penrose1989emperor}.
Rovelli's relational quantum mechanics \cite{rovelli1996relational}
concludes that reality is purely subjective based on an analysis of
various interpretations, thereby offering an explanation and resolution
to the measurement problem. Many other interpretations exist with
their own good merit in explaining measurement. See Ref \cite{drummond2019understanding,zeilinger1996interpretation}
for an overview of interpretations.

\section{Two-photon Interference}

\begin{figure}[h]
\begin{centering}
\includegraphics[width=0.5\linewidth]{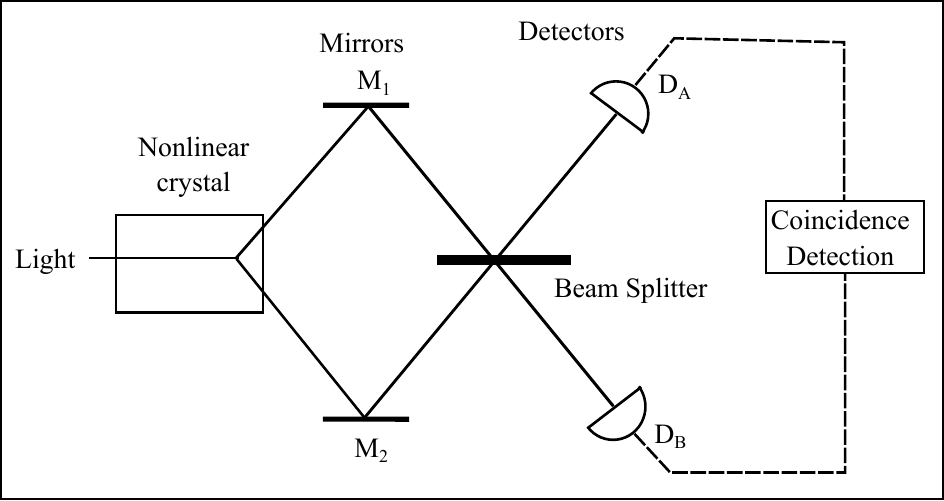}
\par\end{centering}
\caption{\label{fig:HOM-Effect}In the Hong-Ou-Mandel effect, a non-linear
crystal is used to split a photon into two identical photons of equal
wavelength. These photons then travel towards a beam splitter, where
they undergo destructive interference. When the photons are detected
by two detectors using coincidence detection, a zero coincidence count
is obtained. This indicates that the two photons are indistinguishable
from each other.}
\end{figure}

The Hong-Ou-Mandel effect \cite{hong1987measurement,branczyk2017hongoumandel,kim2020hong}
is a two-photon quantum interference effect used in the thought experiment.
Here is a brief overview of the effect:

First, consider a non-linear crystal that generates a pair of identical
photons of equal wavelength through a down-conversion process. These
photons then travel towards two mirrors, $M_{1}$ and $M_{2}$, and
are reflected back towards a beam splitter. Finally, the photons are
detected by two detectors, $D_{A}$ and $D_{B}$, by coincidence detection.
In coincidence detection, a detection is counted only if one photon
is detected at detector $D_{A}$ and the other photon is detected
at detector $D_{B}$ at the same time. Figure {[}\ref{fig:HOM-Effect}{]}
illustrates the setup for the Hong-Ou-Mandel effect.

The two photons that reach the beam splitter can either be transmitted,
reflected, or have one transmitted and the other reflected. Upon reflection,
a photon experiences a phase shift of $\pi/2$. The state of the detectors
can be written as:

\begin{equation}
\vert \phi\rangle =\frac{1}{2}\left(\vert 1_{A}1_{B}\rangle +e^{i\pi/2}\vert 2_{A}0_{B}\rangle +e^{i\pi/2}\vert 0_{A}2_{B}\rangle +e^{i\pi}\vert 1_{A}1_{B}\rangle \right)\label{eq:HOM-1}
\end{equation}

The notation $\vert 0\rangle $, $\vert 1\rangle $, and
$\vert 2\rangle $ represent the number of detections, with
the subscripts denoting the detectors $D_{A}$ and $D_{B}$. Because
the photons are indistinguishable and have equal wavelength, the first
and last terms in the equation can cancel each other out. Thus-

\begin{equation}
\vert \phi\rangle =\frac{e^{i\pi/2}}{2}\left(\vert 2_{A}0_{B}\rangle +\vert 0_{A}2_{B}\rangle \right)\label{eq:HOM-2}
\end{equation}

The equation above shows that the two photons are always detected
together, either at one detector or the other. It is not possible
for each detector to detect a photon, resulting in a zero coincidence
count. This is effectively a way to test the distinguishability between
two photons without directly measuring their states. 

However, if the two photons are not indistinguishable, the first and
last terms in the equation cannot cancel each other out, and all four
possibilities can occur. As a result, from Equation (\ref{eq:HOM-1}),
it can be concluded that half of the total detection events would
be coincidence counts. See Ref \cite{branczyk2017hongoumandel} for
more on Hong-Ou-Mandel effect. 

\section{Experimental Setup}

This section describes the thought experiment. Consider the following
measurement process: choose a beam of circularly polarized light in
the state-

\begin{equation}
\vert \psi\rangle =\frac{1}{\sqrt{2}}\left(\vert R\rangle +\vert L\rangle \right)\label{eq:Measurement-Process-1}
\end{equation}

\begin{figure}[h]
\begin{centering}
\includegraphics[width=0.5\linewidth]{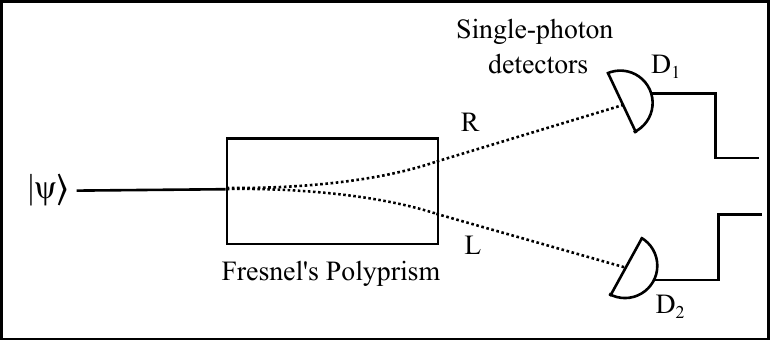}
\par\end{centering}
\caption{\label{fig:Measurement-Process}The figure shows a measurement process
where a beam of circularly polarized light passes through a Fresnel
polyprism, splitting it into right and left circularly polarized light.
The measuring apparatus consists of two single-photon detectors located
at the two beams, denoted as $D_{1}$ and $D_{2}$, respectively.}
\end{figure}

$\vert R\rangle $ and $\vert L\rangle $ indicate right
and left circularly polarized light, respectively. When the beam of
light passes through a Fresnel polyprism \cite{arteaga2019stern},
it splits into two beams of right and left circularly polarizations.
These beams are then detected by two single-photon detectors \cite{eisaman2011invited}
which constitute the measuring apparatus. This is similar to the Stern-Gerlach
experiment with silver atoms \cite{gerlach1922experimentelle}. See
Figure {[}\ref{fig:Measurement-Process}{]}.

The state after measurement is-

\begin{equation}
\vert \psi\rangle =\frac{1}{\sqrt{2}}\left(\vert R\rangle \vert D_{1}\rangle +\vert L\rangle \vert D_{2}\rangle \right)\label{eq:Measurement-Process-2}
\end{equation}

The states $\vert D_{1}\rangle $ and $\vert D_{2}\rangle$
correspond to the detection of a photon at detectors $D_{1}$ and
$D_{2}$, respectively. These detectors are coupled with two single-photon
emitters \cite{eisaman2011invited} $E_{1}$ and $E_{2}$. When a
detector detects a photon, it emits an electrical pulse, triggering
its coupled single-photon emitter to emit a photon. Let the photons
emitted by emitters $E_{1}$ and $E_{2}$ have horizontal and vertical
polarizations, respectively. The state of the system can now be written
as-

\begin{equation}
\vert \psi\rangle =\frac{1}{\sqrt{2}}\left(\vert R\rangle \vert D_{1}\rangle \vert H\rangle +\vert L\rangle \vert D_{2}\rangle \vert V\rangle \right)\label{eq:Measurement-Process-3}
\end{equation}

\begin{figure}
\begin{centering}
\includegraphics[width=0.6\linewidth]{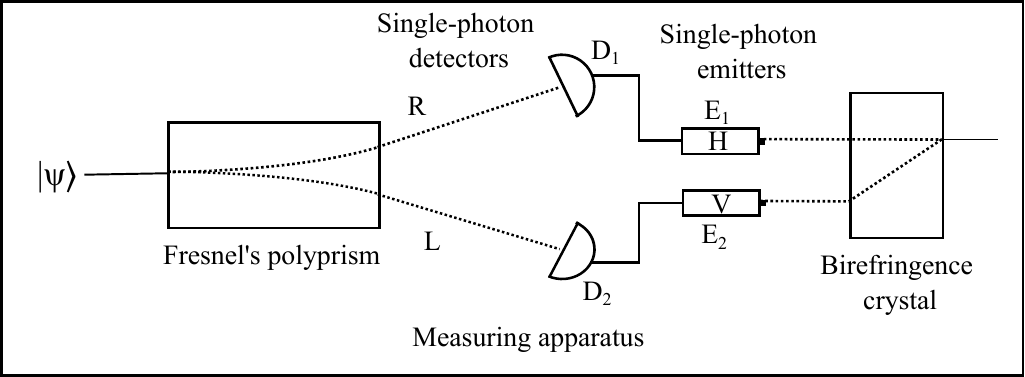}
\par\end{centering}
\caption{\label{fig:Measuring-Apparatus-With-Emitters}After passing through
a Fresnel polyprism, a beam of light splits into right and left circularly
polarized beams. These two beams are then detected at two separate
detectors, which are coupled to two emitters. The emitters generate
two new photons with horizontal and vertical polarizations. The paths
of these photons are merged into a single path using a birefringent
crystal.}

\end{figure}

The states $\vert H\rangle $ and $\vert V\rangle $ indicate
the horizontal and vertical polarization of the emitted photon, respectively.
The two photon paths are then merged into one by a birefringent crystal,
as depicted in Figure {[}\ref{fig:Measuring-Apparatus-With-Emitters}{]}.
In order to observe superposition, it is crucial to ensure indistinguishability
between the different outcomes \cite{zeilinger1999experiment}. It
means there should be no counters or records of any kind. It should
not be possible to determine the outcome of the measurement even in
princple \cite{zeilinger1999experiment}.

\begin{figure}
\includegraphics[width=1\linewidth]{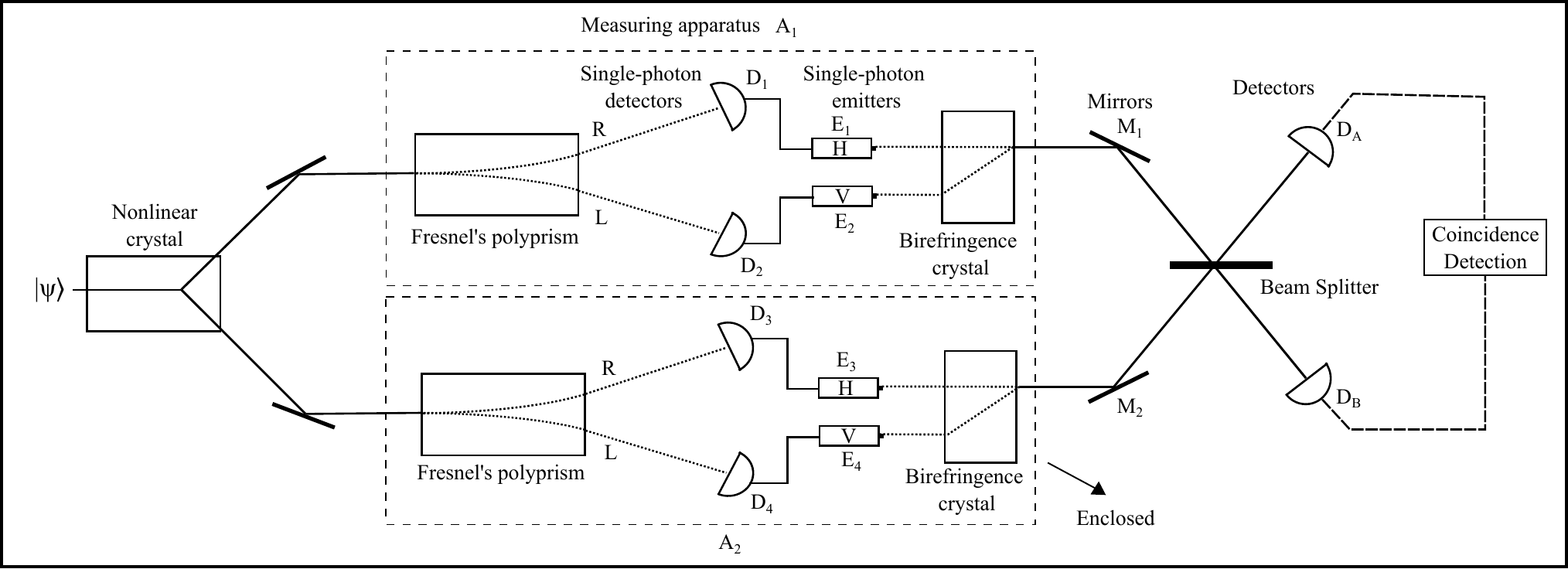}

\caption{\label{fig:Thought-Experiment}Two identical photons produced by a
non-linear crystal are measured by two identical measuring apparatuses.
These apparatuses then generate two new photons, which are sent through
a Hong-Ou-Mandel interferometer setup for coincidence detection. A
zero coincidence count implies the existence of superposition in the
measuring apparatus.}

\end{figure}

Consider a second apparatus that is identical to the one described
earlier, from which another identical photon is obtained. These two
photons are then input into a Hong-Ou-Mandel interferometer setup.
The two identical measuring apparatuses are made to measure two down-converted
photons coming out of a non-linear crystal similar to the original
Hong-Ou-Mandel experiment \cite{hong1987measurement}. See Figure
{[}\ref{fig:Thought-Experiment}{]}.

\section{Discussion}

In the thought experiment, there are two possible outcomes: a zero
coincidence count and a non-zero coincidence count. If the result
is a zero coincidence count, it indicates that the two photons entering
the interferometer are indistinguishable from each other. For this
to occur, each photon can only be in a superposition state, which
suggests that wave function collapse did not occur in the measuring
apparatus. As wave function collapse did not occur, it implies that
the measuring apparatus existed in a superposition state following
the measurement of circularly polarized photons at the beginning.
Thus, measuring apparatus exhibits a superposition state.

If a non-zero coincidence count is observed, it indicates that the
two photons entering the interferometer are not indistinguishable
from each other. Hence, a wave function collapse must have occurred
in the measuring apparatus. In this case, the photon emitted from
the measuring apparatus has an equal probability of being horizontally
or vertically polarized. The system of two photons entering the Hong-Ou-Mandel
interferometer has four possible polarization states.

\begin{equation}
\vert \psi_{1}\psi_{2}\rangle =\frac{1}{2}\left(\vert H_{1}H_{2}\rangle +\vert H_{1}V_{2}\rangle +\vert V_{1}H_{2}\rangle +\vert V_{1}V_{2}\rangle \right)\label{eq:Collapsed-State-Calculation-1}
\end{equation}

$\vert H\rangle $ and $\vert V\rangle $ represent horizontal
and vertical polarization, respectively, and the subscripts $1$ and
$2$ indicate the photons from the first and second measuring apparatus.
In the Hong-Ou-Mandel interferometer, if the two photons have identical
polarization, they will produce a zero coincidence count due to their
indistinguishability. If the two photons have different polarization,
the coincidence count would be half of the total detection events.
A simple calculation shows that one-fourth of the total detection
events are coincidence counts if a wave function collapse does occur.

\begin{equation}
\frac{\textrm{Coincidence Detection Events}}{\textrm{Total Detection Events}}=\frac{0+0.5+0.5+0}{1+1+1+1}=\frac{1}{4}\label{eq:Collapsed-State-Calculation-2}
\end{equation}

\section{Conclusion}

This paper presents a thought experiment that explores the measurement
problem in quantum mechanics, specifically concerning wave function
collapse. It identifies two distinct experimental outcomes based on
it; a zero coincidence count shows superposition existed in the measuring
apparatus while a non-zero coincidence count shows otherwise. In the
thought experiment, using a destructive measurement process preserves
indistinguishability of different states for the detectors as the
original system in superposition is destroyed by the measurement process
itself. Indistinguishbility is crucial for any system to exhibit superposition.
Passing on the superposition from the measuring apparatus to a photon
reduces the time during which the measuring apparatus exists in a
superposition state, thus also help preserve the superposition. Furthermore,
using two-photon interference decreases the likelihood of wave function
collapse occuring due to additional observers required otherwise. 

The standard formulation of quantum mechanics predicts the measuring
apparatus to be in a superposition state when treated as a quantum
system. However, this only holds in ideal conditions where decoherence
effects are negligible and indistinguishability is maintained. The
ideal situation would involve only one object containing the superposition
state at any given time in the thought experiment. However, in practice,
the devices are not perfect and reset after a few nanoseconds, which
may affect the preservation of indistinguishability due to forming
short-lived entangled composite systems. Scaling down the devices
may help reduce these brief intervals and move closer to the ideal
scenario.

This thought experiment provides a framework for discussing the measurement
problem in quantum mechanics. It helps probe the exact conditions
necessary for a wave function collapse to happen. It helps highlight
the importance of indistinguishability to wave function collapse.
Experimental demonstration of the thought experiment would show that
the wave function collapse, if it occurs, does not occur immediately
on measurement. This may help scrutinizing various interpretations.
It is closely related to other experiments which highlight the different
aspects of wave function collapse (Schr{\"o}dinger's cat, Wigner's friend,
Wheeler's delayed choice etc.). Overall, this study contributes to
the exploration of the measurement problem in quantum mechanics and
provides insights into the phenomenon of wave function collapse.

This work has benefited substantially through discussions with \textit{Prof.
Urjit Yajnik}, IIT-Bombay, India. The author has no other relevant
financial or non-financial interests to disclose.

\bibliography{Bibliography}

\end{document}